\RequirePackage{fix-cm}

\documentclass[aps,twocolumn,showpacs,superscriptaddress,tightenlines,floatfix]{revtex4}

\usepackage{amsmath}
\usepackage{amssymb}
\usepackage{graphicx}
\usepackage{epsfig}
\usepackage{color}
\usepackage{multirow}
\usepackage{feynmp}
\usepackage{tikz-feynman}

\begin{document}
\title{Inclusive study of bottomonium production in association with an $\eta$ meson in $e^+e^-$ annihilations near $\Upsilon(5S)$}

\noaffiliation
\affiliation{Aligarh Muslim University, Aligarh 202002}
\affiliation{University of the Basque Country UPV/EHU, 48080 Bilbao}
\affiliation{Beihang University, Beijing 100191}
\affiliation{University of Bonn, 53115 Bonn}
\affiliation{Budker Institute of Nuclear Physics SB RAS, Novosibirsk 630090}
\affiliation{Faculty of Mathematics and Physics, Charles University, 121 16 Prague}
\affiliation{Chiba University, Chiba 263-8522}
\affiliation{Chonnam National University, Kwangju 660-701}
\affiliation{University of Cincinnati, Cincinnati, Ohio 45221}
\affiliation{Deutsches Elektronen--Synchrotron, 22607 Hamburg}
\affiliation{University of Florida, Gainesville, Florida 32611}
\affiliation{Department of Physics, Fu Jen Catholic University, Taipei 24205}
\affiliation{Justus-Liebig-Universit\"at Gie\ss{}en, 35392 Gie\ss{}en}
\affiliation{Gifu University, Gifu 501-1193}
\affiliation{II. Physikalisches Institut, Georg-August-Universit\"at G\"ottingen, 37073 G\"ottingen}
\affiliation{SOKENDAI (The Graduate University for Advanced Studies), Hayama 240-0193}
\affiliation{Gyeongsang National University, Chinju 660-701}
\affiliation{Hanyang University, Seoul 133-791}
\affiliation{University of Hawaii, Honolulu, Hawaii 96822}
\affiliation{High Energy Accelerator Research Organization (KEK), Tsukuba 305-0801}
\affiliation{J-PARC Branch, KEK Theory Center, High Energy Accelerator Research Organization (KEK), Tsukuba 305-0801}
\affiliation{Hiroshima Institute of Technology, Hiroshima 731-5193}
\affiliation{IKERBASQUE, Basque Foundation for Science, 48013 Bilbao}
\affiliation{University of Illinois at Urbana-Champaign, Urbana, Illinois 61801}
\affiliation{Indian Institute of Science Education and Research Mohali, SAS Nagar, 140306}
\affiliation{Indian Institute of Technology Bhubaneswar, Satya Nagar 751007}
\affiliation{Indian Institute of Technology Guwahati, Assam 781039}
\affiliation{Indian Institute of Technology Hyderabad, Telangana 502285}
\affiliation{Indian Institute of Technology Madras, Chennai 600036}
\affiliation{Indiana University, Bloomington, Indiana 47408}
\affiliation{Institute of High Energy Physics, Chinese Academy of Sciences, Beijing 100049}
\affiliation{Institute of High Energy Physics, Vienna 1050}
\affiliation{Institute for High Energy Physics, Protvino 142281}
\affiliation{Institute of Mathematical Sciences, Chennai 600113}
\affiliation{INFN - Sezione di Napoli, 80126 Napoli}
\affiliation{INFN - Sezione di Torino, 10125 Torino}
\affiliation{Advanced Science Research Center, Japan Atomic Energy Agency, Naka 319-1195}
\affiliation{J. Stefan Institute, 1000 Ljubljana}
\affiliation{Kanagawa University, Yokohama 221-8686}
\affiliation{Institut f\"ur Experimentelle Kernphysik, Karlsruher Institut f\"ur Technologie, 76131 Karlsruhe}
\affiliation{Kavli Institute for the Physics and Mathematics of the Universe (WPI), University of Tokyo, Kashiwa 277-8583}
\affiliation{Kennesaw State University, Kennesaw, Georgia 30144}
\affiliation{King Abdulaziz City for Science and Technology, Riyadh 11442}
\affiliation{Department of Physics, Faculty of Science, King Abdulaziz University, Jeddah 21589}
\affiliation{Korea Institute of Science and Technology Information, Daejeon 305-806}
\affiliation{Korea University, Seoul 136-713}
\affiliation{Kyoto University, Kyoto 606-8502}
\affiliation{Kyungpook National University, Daegu 702-701}
\affiliation{\'Ecole Polytechnique F\'ed\'erale de Lausanne (EPFL), Lausanne 1015}
\affiliation{P.N. Lebedev Physical Institute of the Russian Academy of Sciences, Moscow 119991}
\affiliation{Faculty of Mathematics and Physics, University of Ljubljana, 1000 Ljubljana}
\affiliation{Ludwig Maximilians University, 80539 Munich}
\affiliation{Luther College, Decorah, Iowa 52101}
\affiliation{University of Malaya, 50603 Kuala Lumpur}
\affiliation{University of Maribor, 2000 Maribor}
\affiliation{Max-Planck-Institut f\"ur Physik, 80805 M\"unchen}
\affiliation{School of Physics, University of Melbourne, Victoria 3010}
\affiliation{Middle East Technical University, 06531 Ankara}
\affiliation{University of Mississippi, University, Mississippi 38677}
\affiliation{University of Miyazaki, Miyazaki 889-2192}
\affiliation{Moscow Physical Engineering Institute, Moscow 115409}
\affiliation{Moscow Institute of Physics and Technology, Moscow Region 141700}
\affiliation{Graduate School of Science, Nagoya University, Nagoya 464-8602}
\affiliation{Kobayashi-Maskawa Institute, Nagoya University, Nagoya 464-8602}
\affiliation{Nara University of Education, Nara 630-8528}
\affiliation{Nara Women's University, Nara 630-8506}
\affiliation{National Central University, Chung-li 32054}
\affiliation{National United University, Miao Li 36003}
\affiliation{Department of Physics, National Taiwan University, Taipei 10617}
\affiliation{H. Niewodniczanski Institute of Nuclear Physics, Krakow 31-342}
\affiliation{Nippon Dental University, Niigata 951-8580}
\affiliation{Niigata University, Niigata 950-2181}
\affiliation{University of Nova Gorica, 5000 Nova Gorica}
\affiliation{Novosibirsk State University, Novosibirsk 630090}
\affiliation{Osaka City University, Osaka 558-8585}
\affiliation{Osaka University, Osaka 565-0871}
\affiliation{Pacific Northwest National Laboratory, Richland, Washington 99352}
\affiliation{Panjab University, Chandigarh 160014}
\affiliation{Peking University, Beijing 100871}
\affiliation{University of Pittsburgh, Pittsburgh, Pennsylvania 15260}
\affiliation{Punjab Agricultural University, Ludhiana 141004}
\affiliation{Research Center for Electron Photon Science, Tohoku University, Sendai 980-8578}
\affiliation{Research Center for Nuclear Physics, Osaka University, Osaka 567-0047}
\affiliation{Theoretical Research Division, Nishina Center, RIKEN, Saitama 351-0198}
\affiliation{RIKEN BNL Research Center, Upton, New York 11973}
\affiliation{Saga University, Saga 840-8502}
\affiliation{University of Science and Technology of China, Hefei 230026}
\affiliation{Seoul National University, Seoul 151-742}
\affiliation{Shinshu University, Nagano 390-8621}
\affiliation{Showa Pharmaceutical University, Tokyo 194-8543}
\affiliation{Soongsil University, Seoul 156-743}
\affiliation{University of South Carolina, Columbia, South Carolina 29208}
\affiliation{Stefan Meyer Institute for Subatomic Physics, Vienna 1090}
\affiliation{Sungkyunkwan University, Suwon 440-746}
\affiliation{School of Physics, University of Sydney, New South Wales 2006}
\affiliation{Department of Physics, Faculty of Science, University of Tabuk, Tabuk 71451}
\affiliation{Tata Institute of Fundamental Research, Mumbai 400005}
\affiliation{Excellence Cluster Universe, Technische Universit\"at M\"unchen, 85748 Garching}
\affiliation{Department of Physics, Technische Universit\"at M\"unchen, 85748 Garching}
\affiliation{Toho University, Funabashi 274-8510}
\affiliation{Tohoku Gakuin University, Tagajo 985-8537}
\affiliation{Department of Physics, Tohoku University, Sendai 980-8578}
\affiliation{Earthquake Research Institute, University of Tokyo, Tokyo 113-0032}
\affiliation{Department of Physics, University of Tokyo, Tokyo 113-0033}
\affiliation{Tokyo Institute of Technology, Tokyo 152-8550}
\affiliation{Tokyo Metropolitan University, Tokyo 192-0397}
\affiliation{Tokyo University of Agriculture and Technology, Tokyo 184-8588}
\affiliation{University of Torino, 10124 Torino}
\affiliation{Utkal University, Bhubaneswar 751004}
\affiliation{Virginia Polytechnic Institute and State University, Blacksburg, Virginia 24061}
\affiliation{Wayne State University, Detroit, Michigan 48202}
\affiliation{Yamagata University, Yamagata 990-8560}
\affiliation{Yonsei University, Seoul 120-749}
\author{U.~Tamponi}\affiliation{INFN - Sezione di Torino, 10125 Torino}
\author{E.~Guido}\affiliation{INFN - Sezione di Torino, 10125 Torino} 
 \author{R.~Mussa}\affiliation{INFN - Sezione di Torino, 10125 Torino} 

  \author{I.~Adachi}\affiliation{High Energy Accelerator Research Organization (KEK), Tsukuba 305-0801}\affiliation{SOKENDAI (The Graduate University for Advanced Studies), Hayama 240-0193} 
  \author{H.~Aihara}\affiliation{Department of Physics, University of Tokyo, Tokyo 113-0033} 
  \author{S.~Al~Said}\affiliation{Department of Physics, Faculty of Science, University of Tabuk, Tabuk 71451}\affiliation{Department of Physics, Faculty of Science, King Abdulaziz University, Jeddah 21589} 
  \author{D.~M.~Asner}\affiliation{Pacific Northwest National Laboratory, Richland, Washington 99352} 
  \author{H.~Atmacan}\affiliation{University of South Carolina, Columbia, South Carolina 29208} 
  \author{V.~Aulchenko}\affiliation{Budker Institute of Nuclear Physics SB RAS, Novosibirsk 630090}\affiliation{Novosibirsk State University, Novosibirsk 630090} 
  \author{T.~Aushev}\affiliation{Moscow Institute of Physics and Technology, Moscow Region 141700} 
% ...
\collaboration{The Belle Collaboration}





\begin{abstract}
We study bottomonium production in association with an $\eta$ meson in $e^+e^-$ annihilations near the $\Upsilon(5S)$, at a center of mass energy of $\sqrt{s}=10.866\,$GeV. The results are based on the $121.4\,$fb$^{-1}$ data sample collected by the Belle experiment at the asymmetric energy KEKB collider. Only the $\eta$ meson is reconstructed and the missing-mass spectrum of $\eta$ candidates is investigated. We observe the $e^+e^-\to\eta\Upsilon_J(1D)$ process and find evidence for the $e^+e^-\to\eta\Upsilon(2S)$ process, while no significant signals of $\Upsilon(1S)$, $h_b(1P)$, nor $h_b(2P)$ are found. Cross sections for the studied processes are reported. 


\end{abstract}

\maketitle

The treatment of the non-perturbative regime of Quantum Chromo-Dynamics represents one of the major open problems in particle physics \cite{Brambilla:2014}. Quarkonia --- bound states of either $b$ and  $\bar{b}$ or $c$ and $\bar{c}$ quarks --- are regarded as one of the most fertile environments in which new theoretical approaches to this quandary can be tested \cite{Brambilla:2010cs}, thanks to the intrinsic multi-scale nature of their dynamics, which are characterized by the co-existence of hard and soft processes \cite{Brambilla:1999xf}. 
The richness of this sector has been shown by the wave of new discoveries from the BaBar, Belle and CLEO experiments, and then BESIII and LHCb, that challenged the prevailing theoretical models for quarkonium spectra and transitions. Unexpected neutral and charged states have been observed in  both charmonium and bottomonium, together with striking violations of the Okubo-Zweig-Iizuka (OZI) rule \cite{Okubo:1963fa,Zweig:1964jf,Iizuka:1966fk} and Heavy Quark Spin Symmetry (HQSS). These have demonstrated that the light-quark degrees of freedom play a crucial role in the description of spectral properties \cite{Voloshin:2016cgm} and transitions \cite{Segovia:2014mca}. For a recent review of the theoretical models of quarkonia, see Refs. \cite{Esposito:2014rxa,Lebed:2016hpi}. 

The study of transitions that violate HQSS, like those on which this work is
focused, is therefore part of a broader topic of studying exotic
quarkonium-like states. HQSS and the models based on it, like the QCD Multipole Expansion ~\cite{Gottfried:1977gp, Bhanot:1979af, Peskin:1979va, Bhanot:1979vb,  Voloshin:1978hc, Voloshin:1980zf}, have been long considered reliable for describing  hadronic transitions in bottomonium. In this approach, the transitions can be classified into favoured non-spin flipping, like $\Upsilon(nS) \to \pi\pi \Upsilon(mS)$, and disfavoured spin-flipping, like $\Upsilon(nS) \to \eta\, \Upsilon(mS)$, which are suppressed by a factor of $(\Lambda_{\mathrm{QCD}}/m_b)^2$. As a result of this suppression, the small ratio of branching fractions
\[
\mathcal{R}^{\eta S}_{\pi\pi S}(n,m)=
\frac{\mathcal{B}[\Upsilon(nS)\to\eta\Upsilon(mS)]}
{\mathcal{B}[\Upsilon(nS)\to\pi^+\pi^-\Upsilon(mS)]} \approx 10^{-3}
\]
is predicted \cite{Kuang:2006me, Voloshin:2007dx}, providing a simple, sensitive and experimentally accessible test of HQSS. HQSS has been verified at $\Upsilon(2S)$ and $\Upsilon(3S)$, with $\mathcal{R}^{\eta S}_{\pi\pi S}(2,1)=
(1.64\pm0.23)\times10^{-3}$~\cite{He:2008xk, BABAR:2011ab, Tamponi:2012rw} and
$\mathcal{R}^{\eta S}_{\pi\pi S}(3,1)<2.3\times10^{-3}$~\cite{BABAR:2011ab} 
but not at $\Upsilon(4S)$: BaBar unexpectedly observed the HQSS-violating  transition
$\Upsilon(4S)\to\eta\Upsilon(1S)$ with a branching fraction of
$(1.96\pm0.28)\times10^{-4}$, $2.41\pm0.42$ larger than the one for the favoured transition $\Upsilon(4S)\to\pi\pi\Upsilon(1S)$ \cite{Aubert:2008az}. A recent Belle measurement \cite{Guido:2017cts} then confirmed this result. This strong disagreement with the HQSS prediction was explained by the contribution of $B$ meson loops or, equivalently, by the
presence of a four-quark $B\bar{B}$ component within the $\Upsilon(4S)$
wave function~\cite{Meng:2008,Voloshin:2011}. In the case of transitions to spin-singlet states, there is still no evidence of $\Upsilon(4S) \to \pi\pi h_b(1P)$, while the $\Upsilon(4S) \to \eta h_b(1P)$ has been observed recently by Belle to be the largest hadronic transition from the $\Upsilon(4S)$ \cite{Tamponi:2015xzb}, with a branching fraction in agreement with theoretical arguments \cite{Guo:2010ca,Segovia:2015raa} based on various treatments of the light-quark contributions.
At the $\Upsilon(5S)$ energy \cite{Adachi:2012,Mizuk:2012pb}, the $\Upsilon(5S) \to \pi\pi h_b(mP)$ transitions, which were expected to be suppressed by the HQSS violation, have been observed by Belle to be enhanced by the presence of intermediate exotic, four-quark states \cite{Bondar:2012,Krokovny:2013}. Finally, the $\Upsilon(5S) \to \omega \chi_{b1}(1P)$ transition has been observed by Belle to be enhanced with respect to the the $\Upsilon(5S) \to \omega \chi_{b2}(1P)$ \cite{He:2014sqj}, contrary to the HQSS expectation if the $\Upsilon(5S)$ were a pure $b\bar{b}$ state \cite{Guo:2014qra}.

\begin{figure}
\includegraphics[width=.9\linewidth]{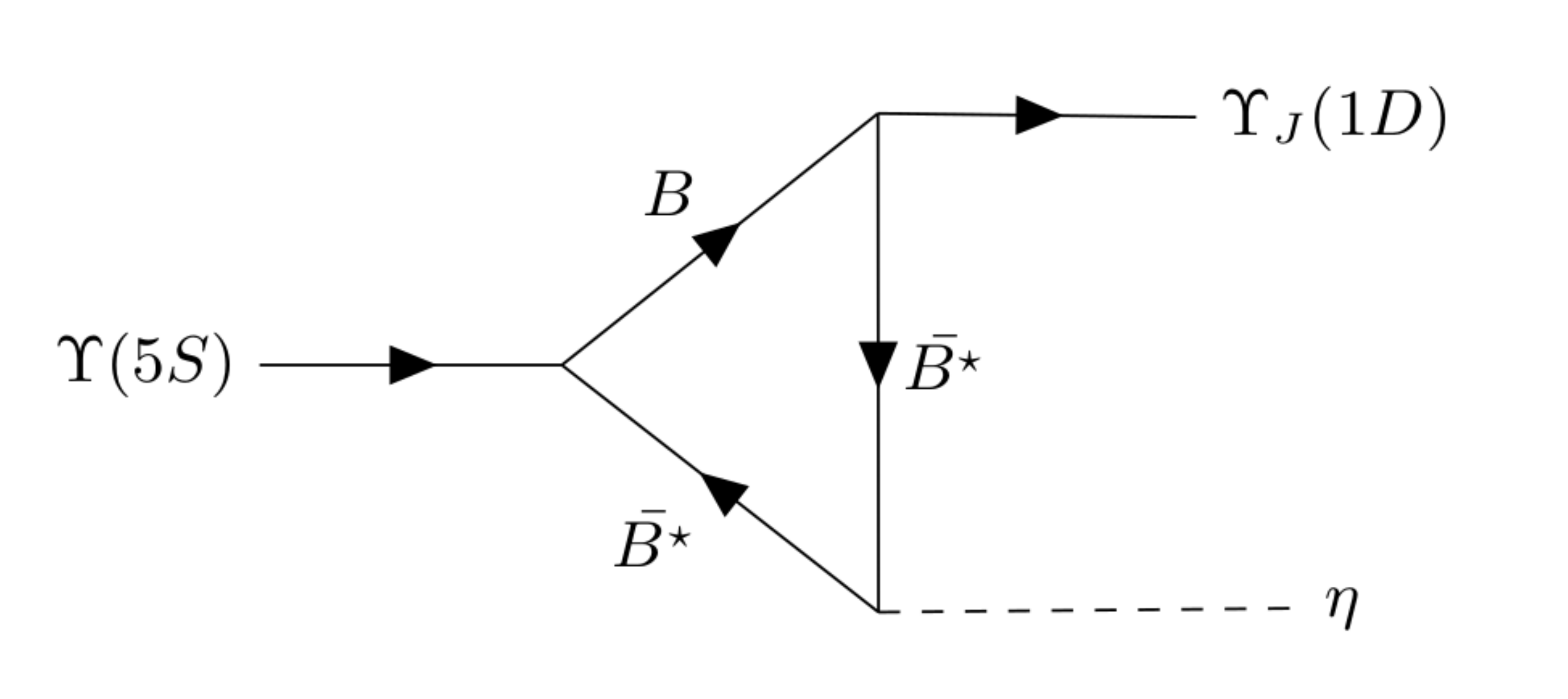}
\caption{Example of triangular $B$ meson loops diagram expected to contribute to the $\Upsilon(5S) \to \eta \Upsilon_J(1D)$ transition, from \cite{Wang:2016qmz}.}
\label{fig:triangularLoop}
\end{figure}

This paper is devoted to the study of one of the missing experimental pieces in the puzzle of the hadronic transitions in bottomonium: the single-$\eta$ emission processes from the $\Upsilon(5S)$ region to $\Upsilon(1S)$, $\Upsilon(2S)$, $\Upsilon_J(1D)$, $h_b(1P)$ and  $h_b(2P)$. The final states with $\Upsilon(1S)$ and $\Upsilon(2S)$ have been studied by theorists using rescattering models \cite{Meng:2008} or by considering intermediate hybrids \cite{Simonov:2008sw}. The predictions are affected by large uncertainties but agree within one order of magnitude with a preliminary result reported by Belle \cite{Krokovny:2012} that was obtained via the exclusive reconstruction of the $\Upsilon(1S, 2S)$ decay into muons. 
In a recent work \cite{Wang:2016qmz}, the case of $\Upsilon(5S) \to \eta \Upsilon_J(1D)$ is analyzed in the context of a rescattering model where the $\Upsilon(5S)$ decays via triangular $B^{(\star)}$ meson loops, as shown in Fig. \ref{fig:triangularLoop}.
The branching fractions are calculated to be of the order of $10^{-3}$, and precise predictions for the contributions due to the three components of the $1D$ triplet are given:
\[
f_1 = \frac{{\cal B}[\Upsilon(5S) \to \eta \Upsilon_1(1D)]}{{\cal B}[\Upsilon(5S) \to \eta \Upsilon(1D_2)]} =  0.68
\]
and  
\[
f_3 = \frac{{\cal B}[\Upsilon(5S) \to \eta \Upsilon_3(1D)]}{{\cal B}[\Upsilon(5S) \to \eta \Upsilon(1D_2)]} =  0.13.
\] 

Our analysis is performed  using the $121.4$ fb$^{-1}$ sample of $e^+e^-$ collisions collected by the Belle experiment nearby the $\Upsilon(5S)$ energy. Following the approach used for the study of $h_b(nP)$ production in $e^+e^-$ collisions at the $\Upsilon(5S)$ \cite{Adachi:2012} and $\Upsilon(4S)$ \cite{Tamponi:2015xzb} energies, we investigate the missing-mass spectrum of $\eta$ mesons in hadronic events. The missing mass is defined as the Lorentz-invariant quantity
$M_{\rm miss}(\eta) c=\sqrt{(P_{e^+e^-}-P_{\eta})^2}$, where $P_{e^+e^-}$
and $P_{\eta}$ are, respectively, the four-momenta of the colliding
$e^+e^-$  pair and the reconstructed $\eta$ meson.

The Belle experiment \cite{Brodzicka:2012} at the KEKB asymmetric $e^+e^-$ collider \cite{kekb,kekb_1,kekb_2} is a $4\pi$ spectrometer optimized for the study of $CP-$violation effects in $B$ meson decays. We highlight here the main characteristics of the apparatus, which is described in detail elsewhere~\cite{Abashian:2000cg}. The tracking of charged particles is provided by four layers of double-sided silicon strip detectors (SVD) and a 50-layer drift chamber (CDC). The energy of photons and electrons is measured by an electromagnetic calorimeter (ECL), while particle identification is obtained by combining the specific ionization measured in the CDC, the time of flight measured by a double layer of plastic scintillators (TOF) and the yield of Cherenkov radiation detected by the Aerogel Cherenkov Counter (ACC). These devices are embedded in a $1.5$T axial magnetic field provided by a cylindrical superconducting solenoid. The iron return yoke of the magnet is instrumented with resistive plate chambers to track and identify muons and $K_L$ mesons. The ECL, which is pivotal for the present measurement, is constructed of CsI(Tl) crystals arranged in a nearly projective geometry to maximize the hermeticity. The central cylindrical barrel covers the polar angle range of $32.2^{\circ} < \theta < 128.7^{\circ}$ while the forward and backward endcap extend the coverage to $\theta = 12^{\circ}$ and  $\theta = 158^{\circ}$, respectively.  The $z$ axis is opposite to the positron beam. 

Studies of the background, optimization of the selection criteria, and estimation of the efficiency are performed using Monte Carlo (MC) samples of the signal processes ({\it signal MC}), and of the $e^+e^- \to B^{(*)}\bar{B^{(*)}}(\pi)$, $e^+e^- \to B^{(*)}_s\bar{B}^{(*)}_s$ and $e^+e^- \to q \bar{q}$ ($q= u,d,s,c$) reactions ({\it generic MC}). The samples are generated using EvtGen \cite{Lange:2001uf}, while the detector response is simulated with GEANT3 \cite{Geant3}. The annihilation of bottomonium into light hadrons, as well as the hadronization of the quarks produced in continuum processes, is simulated by Pythia6~ \cite{Sjostrand:2006za}. The angular distributions of the signal processes are generated assuming the lower angular momentum amplitudes to be dominant.
Separate MC samples are generated for each run period to account for evolution in the detector performance and accelerator conditions.
Each selection criterion is optimized separately, maximizing the figure of merit $F = N_{S}/\sqrt{N_B}$, where $N_{S(B)}$ is the number of signal (background) events passing the selection. To ensure that the selection is independent of the $\eta$ meson momentum, most of the optimization is performed using as signal all the $\eta$ mesons present in the generic MC samples. The signal MC is used only to optimize the suppression of $e^+e^- \to q \bar{q}$ events and to estimate the reconstruction efficiency.

The analysis procedure is similar to the one described in Ref. \cite{Tamponi:2015xzb}, where the process $\Upsilon(4S) \to \eta\, b\bar{b}$ is considered.
An $\eta$ candidate is reconstructed in the $\gamma \gamma$ channel only; the 3$\pi$ modes, both charged and neutral, are not considered due to the low reconstruction efficiency and larger combinatorial background. The $\gamma$ candidates are selected from energy deposits in the ECL not associated with charged tracks. ECL clusters induced by neutral hadrons are suppressed by requiring the shower's transvers-profile radius to be less than $5.1$ cm and the ratio of the energy deposits in a 3$\times$3 and 5$\times$5 crystal matrix around the cluster center to be greater than $0.9$. Since the beam-induced background produces low-energy clusters mostly in the endcap regions, we apply a minimum photon energy  threshold that varies as a function of the cluster polar angle:  $E_{\gamma} > 75 $ MeV in the backward ECL endcap, $E_{\gamma} > 50$ MeV in the backward half of the barrel,  $E_{\gamma} > 60$ MeV in the forward barrel, and $E_{\gamma} > 95 $ MeV in the forward endcap.
The absolute photon energy and the ECL resolution are calibrated by comparing, respectively, the peak position and the widths of three calibration signals, $\pi^0 \to \gamma \gamma$, $\eta \to \gamma\gamma$, and $D^{*0} \to D^{0}\gamma$, in the MC sample and the data \cite{Mizuk:2012pb}. Averaging the results from the different samples, we obtain an energy-scale correction ${\cal F}_{en}(E) = (0.67 \pm 0.25)\%$ at $E_{\gamma} = 0.1$ GeV, that first decreases to $(0.05 \pm 0.23)\%$ at $E_{\gamma} = 0.7$ GeV, and then increases again up to $(0.30 \pm 0.20)\%$  at $E_{\gamma} = 1.4$ GeV. The resolution correction factor decreases smoothly with the photon energy $E_{\gamma}$, from  ${\cal F}_{res}(E) = (25 \pm 10)\%$ at $E_{\gamma} = 0.1$ GeV to $(1 \pm 3)\%$ at $E_{\gamma} = 1.4$ GeV. These are used to calibrate the simulated events. An iterative $\pi^0$-veto procedure removes from the $\eta$-candidate daughter list the photons that are associated with a $\pi^0 \to \gamma \gamma$ decay. Such photons are selected from pairs with an invariant mass $M(\gamma\gamma)$ within $17$ MeV/$c^2$ of the nominal $\pi^0$ mass $m_{\pi^0}$ \cite{Olive:2016xmw}. At each iteration, we remove from the $\eta$ daughter list the photon pair with mass closest to $m_{\pi^0}$, and we update the $\pi^0$ list so that the excluded photons are not used to construct further $\pi^0$ candidates. Finally, we exploit the scalar nature of the $\eta$ to further suppress the combinatorial background, by requiring the photon helicity angle (\textit{i.e.}, the angle $\theta$ between the photon direction and that of the $\Upsilon(5S)$ in the $\eta$ rest frame) to satisfy  $\cos\theta < 0.94$.
The resolution on the $\eta$ invariant mass is $13$ MeV/$c^2$. Candidates with invariant mass  within $26$ MeV/$c^2$ of the nominal $\eta$ mass $m_{\eta}$ \cite{Olive:2016xmw} are selected for the signal sample, while those in the regions $39$ MeV/$c^2$ $<|M(\gamma\gamma) - m_{\eta}| < 52$ MeV/$c^2$ are used as control samples (sidebands). In both cases, we constrain the $\gamma\gamma$ invariant mass to the world-average $\eta$ mass to improve the resolution on the missing mass.
To reduce QED backgrounds $e^+e^- \to (n\gamma) + e^+e^-, \mu^+\mu^-,\tau^+\tau^-$, we apply the Belle standard selection for hadronic events \cite{Abe:2001hj} by requiring each event to have more than two charged tracks pointing towards the primary interaction vertex, a total visible energy greater than $0.2\sqrt{s}$ (where $\sqrt{s}$ is the center-of-mass energy of the $e^+e^-$ pair), a total energy deposit in the ECL between $0.1\sqrt{s}$ and $0.8\sqrt{s}$, and a total momentum balanced along the beam axis.
Continuum $e^+e^- \to q\bar{q}$ events, which are the largest contributor to the background, are characterized by a distinct event topology and are suppressed with the requirement on the ratio of Fox-Wolfram moments $R_2 = H_2/H_0$ \cite{Fox:1978vu} to be less than $0.3$.
\begin{figure}[ht!]
   \includegraphics[width=1.0\linewidth]{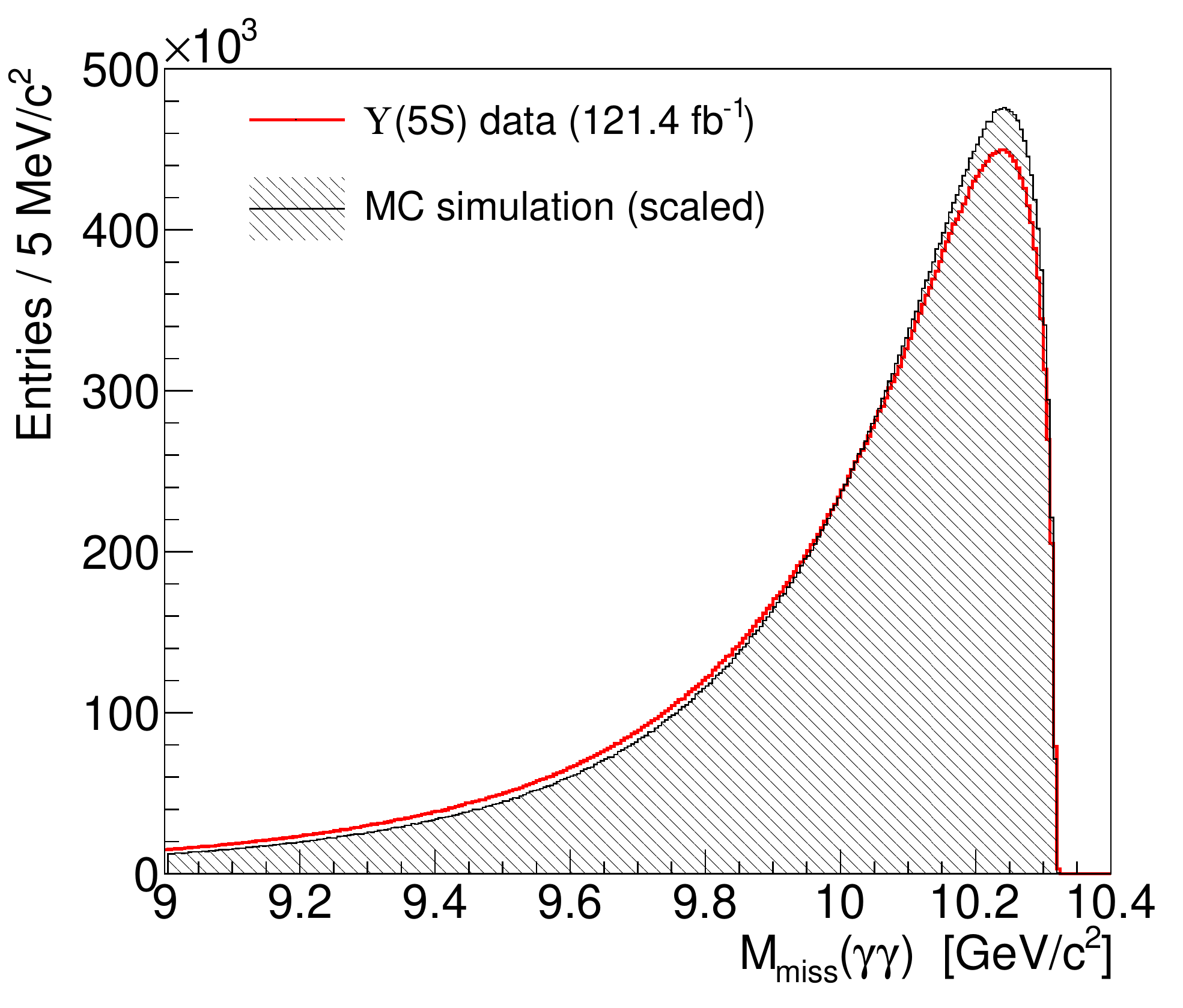}
   \caption{Missing mass of the $\eta$ candidate after the selection. The distribution obtained in the data (red solid histogram) is compared with the MC expectation (black shaded histogram), rescaled by a factor $1.49$. The binning shown here is 50 times larger than the one used in the fitting procedure.}
   \label{fig:rawrecoil}
\end{figure}
Fitting the $M(\gamma\gamma)$ distribution, we estimate the purity of the selected $\eta$ candidates to be $13\%$.
The comparison between the MC simulation and the data is shown in Fig. \ref{fig:rawrecoil}. The MC simulation underestimates the number of events in the $\eta$ invariant mass window by a factor of $1.49$, and does not accurately describe the shape of the distribution observed in the data. We attribute this effect to a non-optimal tuning of the Pythia6 parameters controlling the $SU(3)_{\mathrm{flavour}}$ breaking effects and the production rates of $\eta$  and $\eta^{\prime}$ mesons. 

\begin{figure*}
\begin{center}
   \includegraphics[width=1.\linewidth]{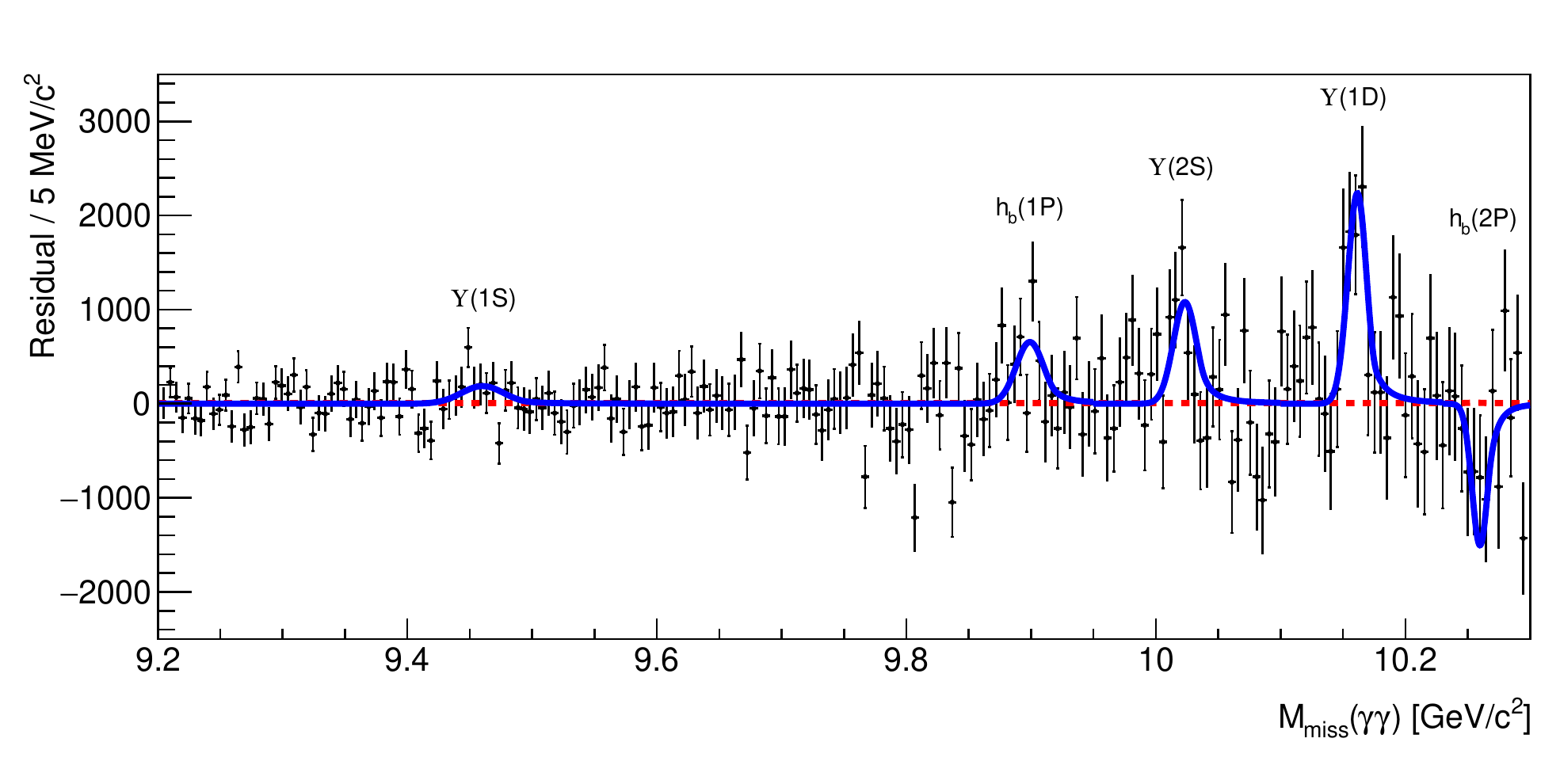}
   \caption{$M_{\mathrm{miss}}(\eta)$ distribution after the subtraction of the fitted background component. The blue solid line shows the signal component of the global fitting function, while the red dashed line represents the background-only component. The binning shown here is 50 times larger than the one used in the fitting procedure.}
   \label{fig:residual_compositepeak_paper}
\end{center}
\end{figure*}
The signal transitions appear as narrow peaks in the $M_{\mathrm{miss}}(\gamma\gamma)$ distribution, whose widths are determined by the resolution on the photon energy reconstruction and the resolution on the beam energy, which is about $5$ MeV. The resulting missing mass resolution decreases almost linearly with $M_{\mathrm{miss}}(\gamma\gamma)$, from $14.1$ MeV/$c^2$ at $M_{\mathrm{miss}}(\gamma\gamma) = m_{\Upsilon(1S)}$ to  $6.3$ MeV/$c^2$ at $M_{\mathrm{miss}}(\gamma\gamma) = m_{h_b(2P)}$. At $M_{\mathrm{miss}}(\gamma\gamma) = m_{\Upsilon(1D)}$, the resolution is $7.6$ MeV/$c^2$.
The $\Upsilon(nS)$ and $h_b(nP)$ signals probability density functions (PDFs) are modeled by Crystal Ball \cite{Gaiser:1986} whose resolutions are fixed to the MC simulation values. The non-Gaussian tail of this PDF captures the effects of the soft initial state radiation (ISR). A simulation method based on the next-to-leading order formula for the ISR emission probability \cite{Benayoun:1999} is used to determine the tail parameters of each signal PDF, which are fixed in the fit. For this calculation, we assume that the energy dependence of the signal cross section is described by a non-relativistic Breit-Wigner function with the parameters of the $\Upsilon(5S)$ resonance \cite{Olive:2016xmw}.
The $\Upsilon_J(1D)$ signal is comprised of three possible states, with unknown fractions and mass splittings. Therefore, we model its PDF as the sum of three separate Crystal Ball functions  ${\cal C}_{J}(m_J)$ with the two scale factors $f_1$ and $f_3$ defined earlier:
\[
{\cal F}_{1D} = \frac{N_{1D}}{1+f_1+f_3}\cdot [{\cal C}_{2}(m_{2}) + f_1 {\cal C}_{1}(m_1) + f_3 {\cal C}_{3}(m_3)],
\]
where $N_{1D}$ is the overall yield of $\Upsilon_J(1D)$, $m_2$ is the $\Upsilon_2(1D)$ mass and the $\Upsilon_{1,3}(1D)$ masses are parametrized as $m_1 = m_2 - \Delta M_{12}$ and  $m_3 = m_2 + \Delta M_{23}$, with $ \Delta M_{ij}$ representing the fine splitting between the $J=1,3$ and  $J = 2$ members of the triplet. To ensure the convergence and stability of the fit, some of the ${\cal F}_{1D}$ parameters are fixed. Theoretical calculations \citep{Eichten:1980mw,Ebert:2002pp,Godfrey:2015dia,Gupta:1982kp,Moxhay:1983vu,Kwong:1988ae,Liu:2011yp,Segovia:2016xqb} and experimental observations \cite{Bonvicini:2004yj,delAmoSanchez:2010kz} suggest that $ \Delta M_{ij} < 10$ MeV/$c^2$; therefore, we fix these to $5$ MeV/$c^2$. Similarly $m_2$ is fixed to the world average value of $10163.7 \pm 1.4$ MeV/$c^2$ \cite{Olive:2016xmw}. The parameters $f_1$, $f_3$ and $N_{1D}$ are allowed to vary. 
The fit is performed in a single region from $9.2$ to $10.3$ GeV/$c^2$ of the binned $M_{\mathrm{miss}}(\gamma\gamma)$ distribution, with a bin width of $0.1$ MeV/$c^2$. The background is modeled with the sum of an ARGUS PDF \cite{Albrecht:1990am} and a seventh-order polynomial. The cut-off parameter of the ARGUS PDF is fixed by the MC simulation, while all the other parameters are allowed to float. The order of the polynomial is chosen to maximize the fit probability.
The result of the fit is shown in Fig. \ref{fig:residual_compositepeak_paper}, where the background PDF has been subtracted to enhance the visibility of the signals. The fit has 17 free parameters ($f_1$ and $f_2$, 10 for the background shape and yield, and 5 signal yields) and a probability of $11\%$. The numerical results are summarized in Table \ref{tab:allFits5S}.
\begin{table}[ht!]
\small
    \caption{Results of the fit of $M_{\mathrm{miss}}(\eta)$. Significance ($\Sigma$), measured signal yield ($N_{\mathrm{meas}}$), and  $\Upsilon_J(1D)$ triplet fractions are reported. The errors on $N_{\mathrm{meas}}$ and $f_J$ are statistical only, while the fit-related systematic uncertainties are taken into account in  the significance estimation.}
    \label{tab:allFits5S}
\begin{center}
\begin{tabular}{lccc}  
\hline    Process                         & $\Sigma$  & $N_{\mathrm{meas}} [10^3]$        \\
\hline 
$e^+e^- \to \eta \Upsilon(1S)$            &  1.5$\sigma$  &$ 1.7  \pm  1.0 $            \\
$e^+e^- \to \eta h_b(1P)$                 &  2.7$\sigma$  &$ 3.9  \pm  1.5$              \\
$e^+e^- \to \eta \Upsilon(2S)$            &  3.3$\sigma$  &$ 5.6  \pm  1.6$           \\
$e^+e^- \to \eta \Upsilon(1D)$            &  5.3$\sigma$  &$ 9.3  \pm  1.8$            \\
$e^+e^- \to \eta h_b(2P)$                 &  $-$ &$ -5.2 \pm  3.6$              \\
\hline
\hline    Fraction                         &    & Fitted value         \\
\hline 
$f_1$                                           &               &$ 0.23 \pm  1.42$             \\
$f_3$                                           &               &$ -0.31 \pm  0.53$             \\
\hline  
\end{tabular}
\end{center}
\end{table}
We observe the $e^+e^- \to \eta \Upsilon_J(1D)$ process and provide evidence for $e^+e^- \to \eta \Upsilon(2S)$. No significant $h_b(nP)$ nor $\Upsilon(1S)$ signals are observed. 

We perform several cross-checks of the fit procedure. First, we verify that the polynomial component has no ripples nor local maxima in the signal regions by studying its first and second derivatives. The fit is then performed on both the MC background-only dataset and a subset of the real data in which the $\gamma\gamma$ pair belongs to the $\eta$ mass sidebands. In both cases, all the signal yields are compatible with zero and the background PDF properly describes the data shape, disfavoring the presence of unaccounted peaking backgrounds. Second, we test a few obvious alternative background models. We replace the ARGUS component with the missing-mass distribution obtained in the background-only MC, then we split the fit range into two sub-ranges above and below $M_{\mathrm{miss}} = 9.8$ GeV/$c^2$, and finally we remove completely the ARGUS component. In all cases, we cannot match the performance of the nominal model without introducing additional free parameters. With the first alternative, we obtain a fit probability of $1\%$  if we increase the polynomial order to 8. With the second alternative we obtain a $10\%$ probability in the upper range, using PDF with an eighth-order polynomial component, and a $5\%$ probability in the lower one  using a third-order polynomial. The third alternative gives a $0.5\%$ fit probability when the polynomial order is increased to 15. We therefore do not regard these as credible alternative models to describe the data.

The visible cross section $\sigma_{v}$ is calculated starting from the fitted yields as $\sigma_{v} = N_{\mathrm{meas}}/\epsilon{\cal B}[\eta \to \gamma \gamma]{\cal L}$, where $N_{\mathrm{meas}}$ is the measured number of signal events, ${\cal L}$ is the integrated luminosity, and $\epsilon$ is the reconstruction efficiency. This quantity can be related to the Born cross section ($\sigma_B$) by de-convolving the ISR effects \cite{Benayoun:1999}:
\[
\sigma_{v}(\sqrt{s}) = \frac{\int_{0}^{\frac{2E_{m}}{\sqrt{s}}} \sigma_{B}(x) W(\sqrt{s}, x) dx}{|1- \Pi|^2}     = \sigma_{B}(\sqrt{s})\frac{1 + \delta_{\mathrm{ISR}}}{|1- \Pi|^2},
\]
\begin{table*}[!t]
\begin{center}
\small
    \caption{Efficiency $\epsilon$, visible cross section $\sigma_v$, ISR correction factor $(1+\delta_{\mathrm{ISR}})$, and Born-level cross section  $\sigma_B$  for the processes considered in this analysis. Upper limits are calculated at $90\%$ confidence level as described in the text.}
    \label{tab:results}
\begin{tabular}{lcccc} 
\hline    Process                          & $\epsilon$ [$\%$]  & $\sigma_v$ [pb]          & $1+\delta_{\mathrm{ISR}}$    &   $\sigma_B$ [pb] \\
\hline 
$e^+e^- \to \eta \Upsilon(1S)$             &  20.1              &$ < 0.34$                 &  $0.644 \pm 0.007$  & $< 0.49$                   \\
$e^+e^- \to \eta h_b(1P)$                  &  22.2              &$ < 0.52$                 &  $0.644 \pm 0.007$  & $< 0.76$                   \\
$e^+e^- \to \eta \Upsilon(2S)$             &  16.5              &$ 0.70 \pm 0.21 \pm 0.12$ &  $0.644 \pm 0.007$  & $1.02 \pm 0.30 \pm 0.17$   \\
$e^+e^- \to \eta \Upsilon_J(1D)$           &  17.2              &$ 1.14 \pm 0.22 \pm 0.15$ &  $0.643 \pm 0.006$  & $1.64 \pm 0.31 \pm 0.21$   \\
$e^+e^- \to \eta h_b(2P)$                  &  16.7              &$ < 0.44    $             &  $0.636 \pm 0.005$  & $<0.64$                    \\
\hline  
\end{tabular}
\end{center}
\end{table*}
where $|1- \Pi|^2 = 0.929$ is the vacuum-polarization factor \cite{Actis:2007fs,He:2014sqj}, $(1 + \delta_{\mathrm{ISR}})$ is the ISR correction factor and $x$ can be interpreted as the fractional energy lost to ISR radiation. The maximum radiated energy is related to the minimum invariant mass of the final hadronic state $M_{\mathrm{min}} = m_{\eta} + m_{(b\bar{b})}$, as $E_m = (s-M^2_{\mathrm{min}})/2\sqrt{s}$. To calculate the value of $(1+\delta_{\mathrm{ISR}})$, we assume that the Born cross section follows a non-relativistic Breit-Wigner shape, and we numerically integrate the expression above. To determine its uncertainty, we repeat the calculation several times, sampling randomly and simultaneously the $\Upsilon(5S)$ parameters and the center of mass energy from Gaussian distributions. The uncertainty on the ISR correction factor is then determined by the spread in the distribution of the $(1 + \delta_{\mathrm{ISR}})$ values. In the process, we assume no correlation among the $\Upsilon(5S)$ mass and width uncertainties.

A summary of the results, including the values of $(1+\delta_{\mathrm{ISR}})$, is presented in Table \ref{tab:results}. To evaluate the upper limits (UL), we use the CL$_s$ modified frequentist method \cite{Read:2002hq} with the profile likelihood ratio as the test statistic. Systematic uncertainties are included by the generation of pseudo-experiments. The significances reported in Table \ref{tab:allFits5S} are evaluated using the asymptotic formulae for the profile likelihood ratio, treating the fit-related systematic uncertainties by including an extra nuisance parameter \cite{Cowan:2010js}. To perform the fits and the statistical analysis, we use the RooFit  \cite{Verkerke:2003ir} and RooStats \cite{Moneta:2010pm} packages.
\begin{table*}[!ht]
\begin{center}
\small
    \caption{Systematic uncertainties, in percentage, in the measurement of the Born-level cross sections of the $e^+e^- \to \eta b\bar{b}$ processes.}
    \label{tab:syst5S}
\begin{tabular}{lccccc} 
\hline Source & $\sigma_B[\eta\Upsilon(1S)]$   & $\sigma_B[\eta\Upsilon(2S)]$  &  $\sigma_B[\eta\Upsilon_J(1D)]$ & $\sigma_B[\eta h_b(1P)]$ & $\sigma_B[\eta h_b(2P)]$  \\
\hline       
\hline       
Luminosity                          &  $\pm 1.4$       &  $\pm 1.4$       &$\pm 1.4$             & $\pm 1.4$   & $\pm 1.4$\\
Reconstruction efficiency           &  $\pm 6.6$       &  $\pm 6.6$       &$\pm 6.6$             & $\pm 6.6$   & $\pm 6.6$\\
$\gamma$ energy calibration         &  $\pm 1.5$       &  $\pm 2.3$       &$\pm 2.8$             & $\pm 2.1$   & $\pm 2.2$\\
Background fit                      &  $\pm 4.0$       &  $\pm 15$        &$\pm 7.1$             & $\pm 4.6$   & $\pm 8.7$\\
Signal model                        &  $\pm 3.2$       &  $\pm 2.5$       &$\pm 8.2$             & $\pm 2.5$   & $\pm 5.5$\\
Radiative correction                &  $\pm 0.6$       &  $\pm 0.6$       &$\pm 0.6$             & $\pm 0.6$   & $\pm 0.8$\\
${\cal B}[\eta \to \gamma\gamma]$   &  $\pm 0.5$       &  $\pm 0.5$       &$\pm 0.5$             & $\pm 0.5$   & $\pm 0.5$\\ 
\hline
Total                               &  $\pm 8.6$       &  $\pm 16.8$      &$\pm 13.1$            &  $\pm 8.8$ &  $\pm 12.5$\\
\hline
\end{tabular}
\end{center}
\end{table*}

We investigate several sources of systematic uncertainty, as summarized in  Table \ref{tab:syst5S}.
The luminosity collected at the $\Upsilon(5S)$ energy has been measured with an uncertainty of  $1.4\%$. The reconstruction efficiency includes several contributions. The photon reconstruction efficiency is known with a  $\pm 2.8\%$ uncertainty per photon, corresponding to $\pm 5.6\%$ per $\eta$,  and has been estimated using $D \to K^{\pm}\pi^{\mp}\pi^{0}$ events. The uncertainty arising from the continuum rejection procedure is estimated  to be  $3.5\%$ by selecting $e^+e^- \to \pi^+\pi^- \Upsilon(2S)$ events and comparing the efficiency of the continuum suppression measured in the data with the one expected from the MC simulation \cite{Adachi:2012}. 
The uncertainty due to the photon energy calibration affects both the signal resolution and the $\eta$ invariant-mass selection. To estimate these effects, we repeat the analysis while varying the calibration factors within their errors.
The background-related uncertainty is obtained by changing simultaneously the polynomial order between 5 and 9, the lower fit-range edge between 9.1 and 9.3 GeV, the upper edge between 10.27 and 10.31 GeV and the bin width between 0.1 and 0.5 MeV. The standard deviation of the distribution of the fit results is then used as the systematic uncertainty. 
The signal model uncertainty is related to the choice of the fixed parameters of the fit. The masses of  $\Upsilon(1S,2S)$ and $h_{b}(1P,2P)$  are varied within their uncertainties and the fit is repeated, obtaining a fluctuation in the signal yields from $2.5\%$ to $5.5\%$, depending on the channel. For the $\Upsilon_J(1D)$, we repeat the fit, changing both the $\Upsilon_J(1D)$ mass within its uncertainties and the values of the splittings between $2$ and $15$ MeV/c${^2}$. To account for possible correlations, we vary all these three parameters independently and simultaneously, repeating the fit under 1960 different configurations. Also, in this case, the standard deviation of the fit result is assumed as a systematic uncertainty. 
The $(1+ \delta_{\mathrm{ISR}})$ factor is calculated with a $\approx 1\%$ precision, according to the channel. Nevertheless, the same parameters that determine the error on $(1+ \delta_{\mathrm{ISR}})$ are also responsible for the uncertainty on the radiative tail in the signal PDF. To estimate the global ISR-related uncertainty, we randomly sample the $\Upsilon(5S)$ parameters and the beam energy as previously described. For each set of parameters, we calculate the ISR correction factor and the signal fit parameters, and then repeat the fit. We find a strong anti-correlation between the fitted signal yields and $1/(1+\delta_{\mathrm{ISR}})$, which means that most of the uncertainty cancels out, leaving only a residual uncertainty of $\approx 0.6\%$. Finally, we include an uncertainty arising from the precision of the world-average value of the $\eta\to\gamma\gamma$ branching fraction \cite{Olive:2016xmw}. 

The behaviour of the hadronic cross section in the $\Upsilon(5S)$ region is not yet entirely understood \cite{Santel:2015qga}. However, assuming that a process proceeds entirely through the $\Upsilon(5S)$ (\textit{i.e.}, there is no continuum contribution, as we assume for the calculation of the ISR correction factor), and that $\sigma[e^+e^- \to \Upsilon(5S)] =  \sigma[e^+e^- \to b \bar{b}] = (0.340 \pm 0.016)$ nb \cite{Esen:2012yz}, an estimation of the branching fraction can be obtained from the visible cross section with the formula
$
{\cal B}[\Upsilon(5S) \to \eta X] = \sigma_v[e^+e^- \to \eta X] / \sigma [e^+e^- \to b \bar{b}].
$
Under these assumptions, we calculate the branching fraction ${\cal B}[\Upsilon(5S) \to \eta \Upsilon_J(1D)] = (4.82 \pm 0.92 \pm 0.67) \times 10^{-3}$. Theoretical calculations that account for the effect of virtual B meson loops are in agreement with our result  \cite{Wang:2016qmz}. 

\begin{figure}
   \includegraphics[width=1.\linewidth]{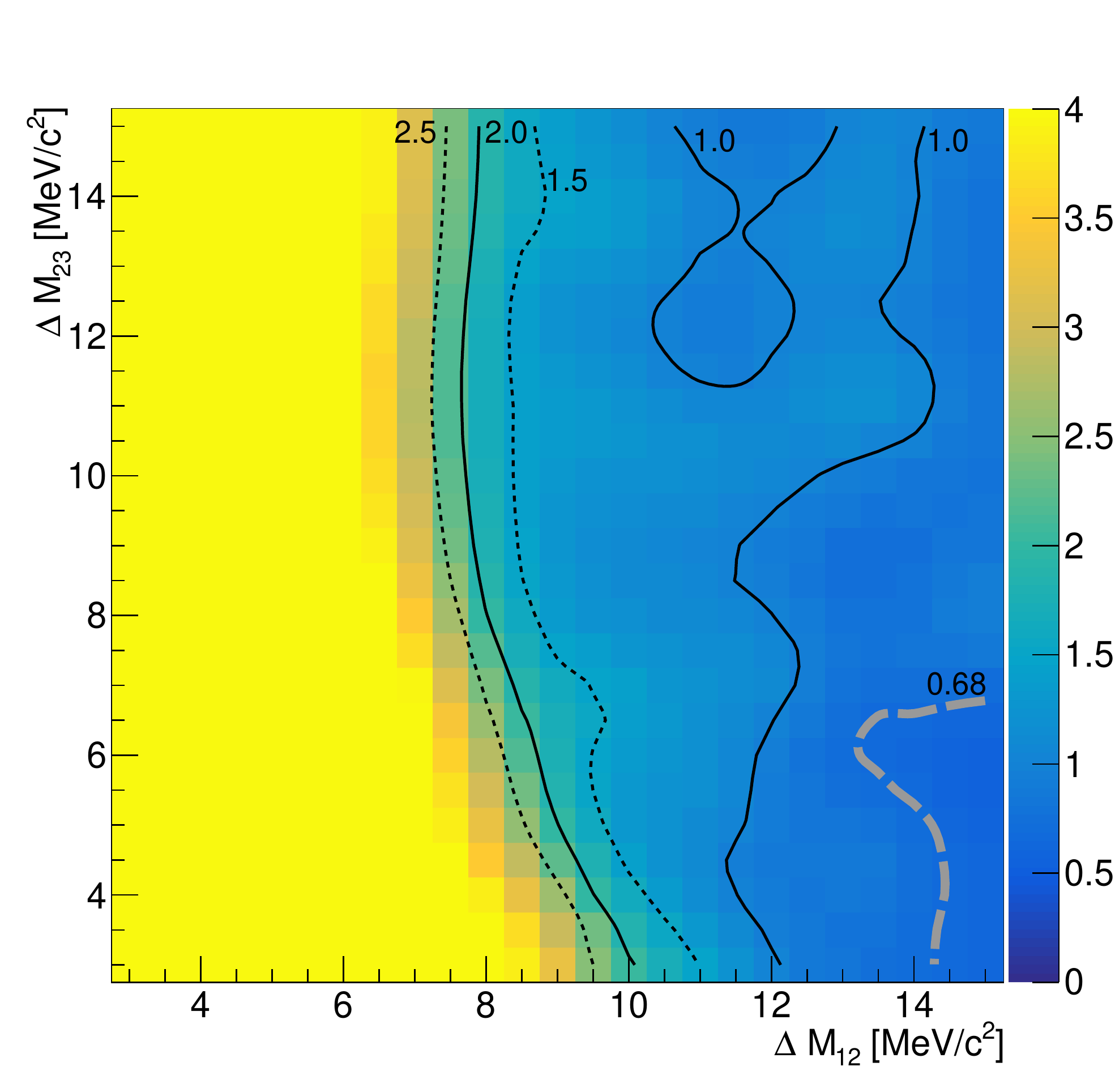}
   \caption{$90\%$ C.L. upper limit on $f_1$, as function of the chosen $\Upsilon_J(1D)$ fine splitting values. The black lines represent the curves at fixed values of the UL in steps of 0.5. The corresponding UL value is reported next to each line. Dashed and solid line styles are alternated for clarity. The thick gray dashed line demarcates the region excluded by the theoretically-favored value $f_1 = 0.68$ \cite{Wang:2016qmz}.}
   \label{fig:upperLimit_1D1}
\end{figure}

\begin{figure}
   \includegraphics[width=1.\linewidth]{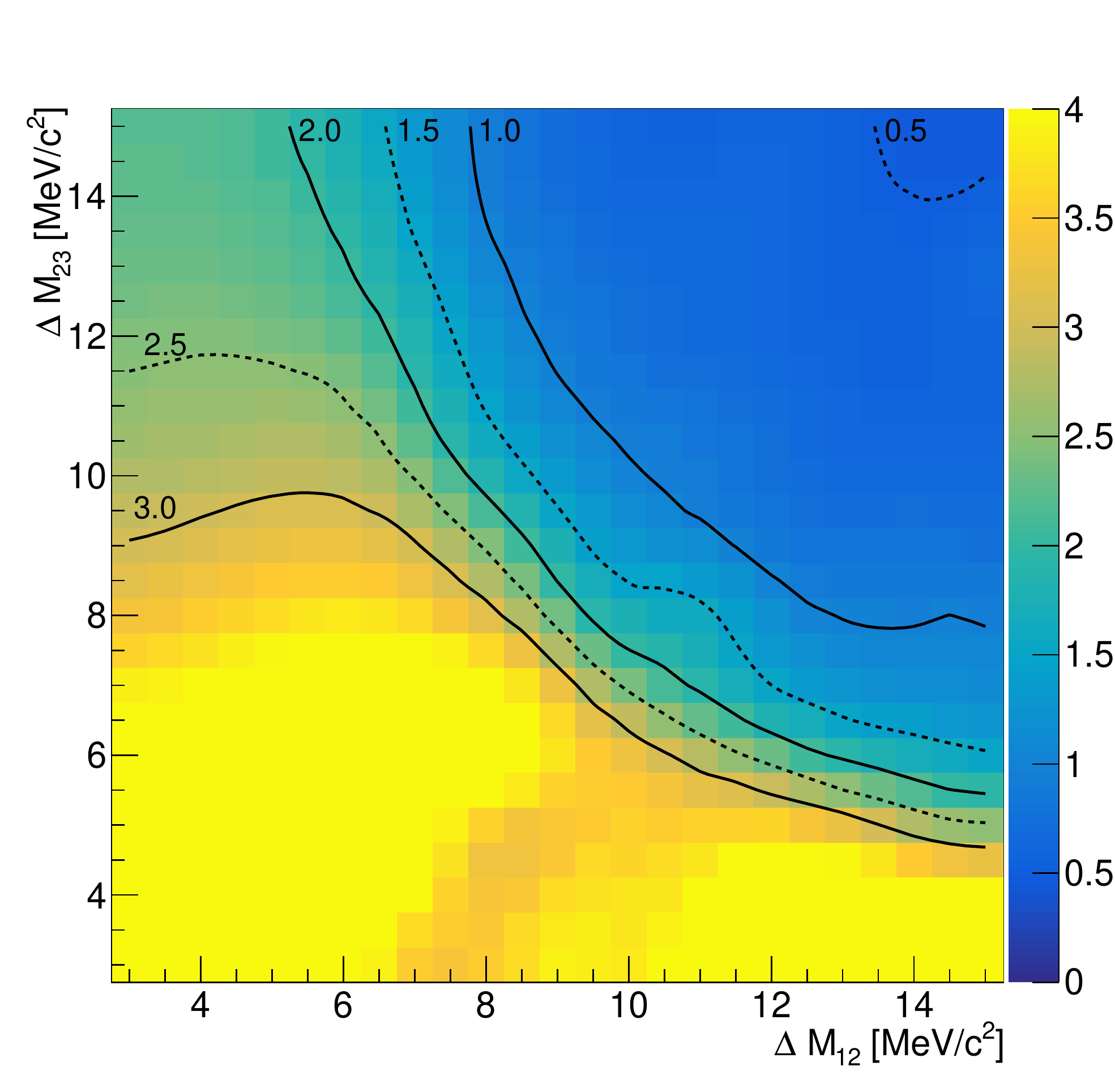}
   \caption{$90\%$ C.L. upper limit on $f_3$, as function of the chosen $\Upsilon_J(1D)$ fine splitting values. The black lines represent the curves at fixed values of the UL in steps of 0.5. The corresponding UL value is reported next to each line. Dashed and solid line styles are alternated for clarity.}
   \label{fig:upperLimit_1D3}
\end{figure}

Our measurements of $f_1$ and $f_3$, the fraction of transitions to $\Upsilon_J(1D)$ that produce the $J=1$ and $J=3$ members of the triplet, respectively, with $\Delta M_{ij} = 5$ MeV/$c^2$, are both compatible with 0. We repeat the fit for other values of the fine splittings in the favoured range of $3$ to $15$ MeV/$c^2$ and, again, do not find significant signals of $J=1$ or $J=3$ states. Possible explanations are that either $\Delta M_{12}$ and $\Delta M_{23}$ are comparable within our experimental resolution, as previous analyses and theoretical predictions suggest, or the $\eta$ transition preferentially produces only one member of the triplet, or both.  We set $90\%$ confidence level (C.L.) upper limits on $f_1$ and $f_3$ as function of $\Delta M_{12}$ and $\Delta M_{23}$, as shown in Fig. \ref{fig:upperLimit_1D1} and \ref{fig:upperLimit_1D3}.
The predictions \cite{Wang:2016qmz} for $f_1$, namely $f_1 = 0.65$, exclude the region where $ \Delta M_{23} \lesssim 7$ MeV/$c^2$ and $ \Delta M_{12} \gtrsim 14$ MeV/$c^2$ (Fig. \ref{fig:upperLimit_1D1}), while the predictions for $f_3$ provide no constraint on either quantity (Fig. \ref{fig:upperLimit_1D3}).

In summary, we report here the first observation of the process $e^+e^- \to \eta\, \Upsilon_J(1D)$ and the first search for  $e^+e^- \to \eta h_b(1P, 2P)$ in the vicinity of the $\Upsilon(5S)$ resonance. The measured visible cross section at $\sqrt{s} = 10.865$ GeV for the former process is $\sigma_{v}[e^+e^- \to \eta \Upsilon_J(1D)] = (1.14 \pm 0.22 \pm 0.15)$ pb. Taking into account the radiative corrections, we measure the Born-level cross section $\sigma_{B}[e^+e^- \to \eta \Upsilon_J(1D)] = (1.64 \pm 0.31 \pm 0.21)$ pb. We also find evidence for the process  $e^+e^- \to \eta \Upsilon(2S)$ and we measure the cross section $\sigma_{v}[e^+e^- \to \eta \Upsilon(2S)] = (0.70 \pm 0.21 \pm 0.12)$ pb, corresponding to $\sigma_{B}[e^+e^- \to \eta \Upsilon(2S)] = (1.02 \pm 0.30 \pm 0.17)$ pb. We do not have significant evidence of  $e^+e^- \to \eta h_b(1P, 2P)$ nor $e^+e^- \to \eta \Upsilon(1S)$. A much larger statistics data set, like the one obtainable with the Belle~II experiment \cite{Abe:2010gxa}, is needed to perform such measurement.
We do not have direct evidence of the presence of the three states of the $\Upsilon_J(1D)$ triplet, and we derive $90\%$ CL upper limits on the fraction of the $J=1$ and $J=3$ state with respect to the $J=2$ state. Our results for the $e^+e^- \to \eta\Upsilon(nS)$  process agrees with a preliminary Belle study in which the exclusive reconstruction of $\Upsilon(1S, 2S)$ into lepton pairs was used~\cite{Krokovny:2012}. 

We thank the KEKB group for the excellent operation of the
accelerator; the KEK cryogenics group for the efficient
operation of the solenoid; and the KEK computer group,
the National Institute of Informatics, and the 
PNNL/EMSL computing group for valuable computing
and SINET5 network support.  We acknowledge support from
the Ministry of Education, Culture, Sports, Science, and
Technology (MEXT) of Japan, the Japan Society for the 
Promotion of Science (JSPS), and the Tau-Lepton Physics 
Research Center of Nagoya University; 
the Australian Research Council;
Austrian Science Fund under Grant No. P 26794-N20;
the National Natural Science Foundation of China under Contracts 
No. 10575109, No. 10775142, No. 10875115, No. 11175187, No. 11475187, 
No. 11521505 and No. 11575017;
the Chinese Academy of Science Center for Excellence in Particle Physics; 
the Ministry of Education, Youth and Sports of the Czech
Republic under Contract No. LTT17020;
the Carl Zeiss Foundation, the Deutsche Forschungsgemeinschaft, the
Excellence Cluster Universe, and the VolkswagenStiftung;
the Department of Science and Technology of India; 
the Istituto Nazionale di Fisica Nucleare of Italy; 
the WCU program of the Ministry of Education, National Research Foundation (NRF)
of Korea Grants No. 2011-0029457, No. 2012-0008143,
No. 2014R1A2A2A01005286,\\
No. 2014R1A2A2A01002734, \\
No. 2015R1A2A2A01003280,
No. 2015H1A2A1033649, 
No. 2016R1D1A1B01010135, 
No. 2016K1A3A7A09005603, 
No. 2016K1A3A7A09005604, 
No. 2016R1D1A1B02012900,
No. 2016K1A3A7A09005606, \\
No. NRF-2013K1A3A7A06056592;
the Brain Korea 21-Plus program, Radiation Science Research Institute, Foreign Large-size Research Facility Application Supporting project and the Global Science Experimental Data Hub Center of the Korea Institute of Science and Technology Information;
the Polish Ministry of Science and Higher Education and 
the National Science Center;
the Ministry of Education and Science of the
Russian Federation under contract 14.W03.31.0026
the Slovenian Research Agency;
Ikerbasque, Basque Foundation for Science and
the Euskal Herriko Unibertsitatea (UPV/EHU) under program UFI 11/55 (Spain);
the Swiss National Science Foundation; 
the Ministry of Education and the Ministry of Science and Technology of Taiwan;
and the U.S.\ Department of Energy and the National Science Foundation.


\begin{thebibliography}{}


\bibitem{Brambilla:2014}
N. Brambilla {\it et al.}, 
Eur. Phys. J. C {\bf 74}, 2981 (2014).

\bibitem{Brambilla:2010cs} 
  N.~Brambilla {\it et al.},
  Eur.\ Phys.\ J.\ C {\bf 71}, 1534 (2011).
  
\bibitem{Brambilla:1999xf}
  N.~Brambilla, A.~Pineda, J.~Soto and A.~Vairo,
  Nucl.\ Phys.\ B {\bf 566}, 275 (2000).
  
  
  
\bibitem{Okubo:1963fa} 
  S.~Okubo,
  Phys.\ Lett.\  {\bf 5}, 165 (1963).

\bibitem{Zweig:1964jf} 
  G.~Zweig,
  Developments in the Quark Theory of Hadrons, Volume 1. Edited by D. Lichtenberg and S. Rosen. pp. 22-101 (1964).
  
\bibitem{Iizuka:1966fk} 
  J.~Iizuka,
  Prog.\ Theor.\ Phys.\ Suppl.\  {\bf 37}, 21 (1966).
  
  
\bibitem{Voloshin:2016cgm}
  M.~B.~Voloshin,
  Phys.\ Rev.\ D {\bf 93},  074011 (2016).
  
 
\bibitem{Segovia:2014mca}
  J.~Segovia, D.~R.~Entem and F.~Fern\`andez,
  Phys.\ Rev.\ D {\bf 91}, 014002 (2015).

  

\bibitem{Esposito:2014rxa}
  A.~Esposito, A.~L.~Guerrieri, F.~Piccinini, A.~Pilloni and A.~D.~Polosa,
  Int.\ J.\ Mod.\ Phys.\ A {\bf 30}, 1530002 (2015).
  
\bibitem{Lebed:2016hpi}
  R.~F.~Lebed, R.~E.~Mitchell and E.~S.~Swanson,
  Prog.\ Part.\ Nucl.\ Phys.\  {\bf 93}, 143 (2017).




\bibitem{Gottfried:1977gp} 
  K.~Gottfried,
  Phys.\ Rev.\ Lett.\  {\bf 40}, 598 (1978).

\bibitem{Bhanot:1979af} 
  G.~Bhanot, W.~Fischler and S.~Rudaz,
  Nucl.\ Phys.\ B {\bf 155}, 208 (1979).


\bibitem{Peskin:1979va} 
  M.~E.~Peskin,
  Nucl.\ Phys.\ B {\bf 156}, 365 (1979).

\bibitem{Bhanot:1979vb} 
  G.~Bhanot and M.~E.~Peskin,
  Nucl.\ Phys.\ B {\bf 156}, 391 (1979).

\bibitem{Voloshin:1978hc} 
  M.~B.~Voloshin,
  Nucl.\ Phys.\ B {\bf 154}, 365 (1979).

\bibitem{Voloshin:1980zf} 
  M.~B.~Voloshin and V.~I.~Zakharov,
  Phys.\ Rev.\ Lett.\  {\bf 45}, 688 (1980).
  
  \bibitem{Kuang:2006me}
  Y.~-P.~Kuang,
  Front.\ Phys.\ China {\bf 1}, 19 (2006).

\bibitem{Voloshin:2007dx}
  M.~B.~Voloshin,
  Prog.\ Part.\ Nucl.\ Phys.\  {\bf 61}, 455 (2008).


\bibitem{He:2008xk}
  Q.~He {\it et al.}  [CLEO Collaboration],
  Phys.\ Rev.\ Lett.\  {\bf 101}, 192001 (2008).

\bibitem{BABAR:2011ab}
  J.~P.~Lees {\it et al.}  [BaBar Collaboration],
  Phys.\ Rev.\ D {\bf 84}, 092003 (2011).

\bibitem{Tamponi:2012rw}
  U.~Tamponi {\it et al.}  [Belle Collaboration],
  Phys.\ Rev.\ D {\bf 87}, 011104 (2013).
  
  \bibitem{Aubert:2008az}
  B.~Aubert {\it et al.}  [BaBar Collaboration],
  Phys.\ Rev.\ D {\bf 78},  112002 (2008).
 
 
\bibitem{Guido:2017cts}
  E.~Guido {\it et al.} [Belle Collaboration],
  Phys.\ Rev.\ D {\bf 96},  052005 (2017).
  
\bibitem{Meng:2008}
  C.~Meng and K.~T.~Chao,
  Phys.\ Rev.\ D {\bf 78}, 074001 (2008).

\bibitem{Voloshin:2011}  
 M.B.~Voloshin,
 Mod.\ Phis.\ Lett.\ A {\bf 26}, 778 (2011).
 
 
\bibitem{Tamponi:2015xzb}
  U.~Tamponi {\it et al.} [Belle Collaboration],
  Phys.\ Rev.\ Lett.\  {\bf 115}, no.14,  142001 (2015).
  
\bibitem{Guo:2010ca}
  F.~-K.~Guo, C.~Hanhart and U.~-G.~Meissner,
  Phys.\ Rev.\ Lett.\  {\bf 105}, 162001 (2010).
 
\bibitem{Segovia:2015raa}
  J.~Segovia, F.~Fernandez and D.~R.~Entem,
  Few Body Syst.\  {\bf 57} no.4, 275 (2016).
  
  
\bibitem{Adachi:2012}
 I.~Adachi {\it et al.}  [Belle Collaboration],
 Phys.\ Rev.\ Lett.\ {\bf 108}, 032001 (2012). 

\bibitem{Mizuk:2012pb} 
  R.~Mizuk {\it et al.}  [Belle Collaboration],
  Phys.\ Rev.\ Lett.\  {\bf 109}, 232002 (2012).
  
  \bibitem{Bondar:2012}
 A.~Bondar {\it et al.}  [Belle Collaboration],
 Phys.\ Rev.\ Lett.\ {\bf 108}, 122001 (2012).

\bibitem{Krokovny:2013}
 P.~Krokovny {\it et al.}  [Belle Collaboration],
  Phys.\ Rev.\ D {\bf 88}, 052016 (2013). 
 
 
 
\bibitem{He:2014sqj} 
  X.~H.~He {\it et al.} [Belle Collaboration],
  Phys.\ Rev.\ Lett.\  {\bf 113}, no. 14, 142001 (2014).
  
\bibitem{Guo:2014qra} 
  F.~K.~Guo, U.~G.~Meißner and C.~P.~Shen,
  Phys.\ Lett.\ B {\bf 738}, 172 (2014).

  
  
  

  
\bibitem{Simonov:2008sw}
  Y.~A.~Simonov and A.~I.~Veselov,
  Phys.\ Lett.\ B {\bf 673}, 211 (2009).
  
  \bibitem{Krokovny:2012}
 P.~Krokovny, talk given at Rencontres de Phisique de la Vallee d'Aoste, La Thuile, Italy (2012).
  
  
\bibitem{Wang:2016qmz} 
  B.~Wang, D.~Y.~Chen and X.~Liu,
  Phys.\ Rev.\ D {\bf 94}, 094039 (2016).
  

\bibitem{Brodzicka:2012}
 J.~Brodzicka  {\it et al.} [for the Belle Collaboration],
 Prog.\ Theor.\ Exp.\ Phys.\  04D001 (2012).
 
 
\bibitem{kekb}
S.~Kurokawa, E.~Kikutani,
Nucl. Instrum. Meth.\ A {\bf 499}, 1 (2003),
and other papers included in this volume.

\bibitem{kekb_1}
T.~Abe {\it et al},
Prog. Theor. Exp. Phys., 03A001 (2013).

\bibitem{kekb_2}
T.~Abe {\it et al},
Prog. Theor. Exp. Phys., 03A006 (2013),
and the other papers included in Prog. Theor. Exp. Phys., Volume 2013 Issue 3 (March 2013). 
 
\bibitem{Abashian:2000cg}
  A.~Abashian {\it et al.},
  Nucl.\ Instrum.\ Meth.\ A {\bf 479}, 117 (2002).



\bibitem{Lange:2001uf}
  D.~J.~Lange,
  Nucl.\ Instrum.\ Meth.\ A {\bf 462}, 152 (2001).
 

\bibitem{Geant3}
  R. Brun {\it et al.}, GEANT3.21, CERN Report DD/EE/84-1 (1984).


\bibitem{Sjostrand:2006za}
  T.~Sjostrand, S.~Mrenna and P.~Z.~Skands,
  JHEP {\bf 0605}, 026 (2006).

\bibitem{Olive:2016xmw}
  C.~Patrignani {\it et al.} [Particle Data Group],
  Chin.\ Phys.\ C {\bf 40}, 100001 (2016).
  
\bibitem{Abe:2001hj}
  K.~Abe {\it et al.}  [Belle Collaboration],
  Phys.\ Rev.\ D {\bf 64} 072001 (2001).

\bibitem{Fox:1978vu}
  G.~C.~Fox and S.~Wolfram,
  Phys.\ Rev.\ Lett.\  {\bf 41}, 1581 (1978).
  
 



  
  \bibitem{Gaiser:1986}
  J.E.~Gaiser {\it et al.}, 
  Phys.\ Rev.\ D {\bf 34}, 711 (1986).
  
  \bibitem{Benayoun:1999}
  M.~Benayoun, S.I.~Eidelman, V.N.~Ivanchenko and Z.K.~Silagadze,
  Mod.\ Phys.\ Lett.\ A {\bf 14}, 2605 (1999).
  
  


\bibitem{Eichten:1980mw}
  E.~Eichten and F.~Feinberg,
  Phys.\ Rev.\ D {\bf 23}, 2724 (1981).
  
\bibitem{Ebert:2002pp}
  D.~Ebert, R.~N.~Faustov and V.~O.~Galkin,
  Phys.\ Rev.\ D {\bf 67}, 014027 (2003).
  
\bibitem{Godfrey:2015dia}
  S.~Godfrey and K.~Moats,
  Phys.\ Rev.\ D {\bf 92}, 054034 (2015).
  
\bibitem{Gupta:1982kp}
  S.~N.~Gupta, S.~F.~Radford and W.~W.~Repko,
  Phys.\ Rev.\ D {\bf 26}, 3305 (1982).
  
  
\bibitem{Moxhay:1983vu}
  P.~Moxhay and J.~L.~Rosner,
  Phys.\ Rev.\ D {\bf 28}, 1132 (1983).

\bibitem{Kwong:1988ae}
  W.~Kwong and J.~L.~Rosner,
  Phys.\ Rev.\ D {\bf 38}, 279 (1988).
  
\bibitem{Liu:2011yp}
  J.~F.~Liu and G.~J.~Ding,
  Eur.\ Phys.\ J.\ C {\bf 72}, 1981 (2012).
  
\bibitem{Segovia:2016xqb}
  J.~Segovia, P.~G.~Ortega, D.~R.~Entem and F.~Fernández,
  Phys.\ Rev.\ D {\bf 93}, 074027 (2016).

\bibitem{Bonvicini:2004yj}
  G.~Bonvicini {\it et al.} [CLEO Collaboration],
  Phys.\ Rev.\ D {\bf 70}, 032001 (2004).
  
\bibitem{delAmoSanchez:2010kz}
  P.~del Amo Sanchez {\it et al.} [BaBar Collaboration],
  Phys.\ Rev.\ D {\bf 82} (2010) 111102
  
\bibitem{Albrecht:1990am}
  H.~Albrecht {\it et al.} [ARGUS Collaboration],
  Phys.\ Lett.\ B {\bf 241}, 278 (1990).

\bibitem{Actis:2007fs}
  S.~Actis, M.~Czakon, J.~Gluza and T.~Riemann,
  Phys.\ Rev.\ Lett.\  {\bf 100}, 131602 (2008).




\bibitem{Read:2002hq}
  A.~L.~Read,
  J.\ Phys.\ G {\bf 28}, 2693 (2002).
  
  
\bibitem{Cowan:2010js} 
  G.~Cowan, K.~Cranmer, E.~Gross and O.~Vitells,
  Eur.\ Phys.\ J.\ C {\bf 71}, 1554 (2011). 
  Erratum: [Eur.\ Phys.\ J.\ C {\bf 73}, 2501 (2013)].
  
\bibitem{Verkerke:2003ir} 
  W.~Verkerke and D.~P.~Kirkby,
  eConf C {\bf 0303241}, MOLT007 (2003).
  
\bibitem{Moneta:2010pm}
  L.~Moneta {\it et al.},
  PoS ACAT {\bf 2010}, 057 (2010).
  

  
  
  
  
  

\bibitem{Santel:2015qga} 
  D.~Santel {\it et al.} [Belle Collaboration],
  Phys.\ Rev.\ D {\bf 93}, 011101 (2016).
  
  
\bibitem{Esen:2012yz}
  S.~Esen {\it et al.} [Belle Collaboration],
  Phys.\ Rev.\ D {\bf 87}, 031101 (2013).

 
  
\bibitem{Abe:2010gxa}
  T.~Abe {\it et al.} [Belle-II Collaboration],
  arXiv:1011.0352.


  
  

\end{thebibliography}
\end{document}